\UseRawInputEncoding
\documentclass[aps,prl,reprint,floatfix,amsmath,amssymb]{revtex4-1}

\usepackage{epsfig}
\usepackage{graphicx}
\usepackage{dcolumn}
\usepackage{bm}
\usepackage[colorlinks=true,citecolor=blue, dvipdfm]{hyperref}
\usepackage{amssymb}

\begin{document}
\title{Complete universal scaling in first-order phase transitions}

\author{Fan Zhong}
\affiliation{School of Physics and State Key Laboratory of Optoelectronic Materials and
Technologies, Sun Yat-sen University, Guangzhou 510275, People's Republic of China}

\date{\today}

\begin{abstract}
 Phase transitions and critical phenomena are among the most intriguing phenomena in nature and society. They are classified as first-order phase transitions (FOPTs) and continuous ones. While the latter show marvelous phenomena of scaling and universality, whether the former behaves similarly is a long-standing controversial issue. Here we definitely demonstrate complete universal scaling in field driven FOPTs for Langevin equations in both zero and two spatial dimensions by rescaling all parameters and subtracting extra contributions with singular dimensions from an effective temperature and a special field according to an effective theory that possesses a fixed point of a seemingly un-physical imaginary value. This offers a perspective different from the usual nucleation and growth but conforming to continuous phase transitions to study FOPTs.
\end{abstract}

\maketitle
Phase transitions and critical phenomena are among the most intriguing phenomena in nature and society. According to a modern classification~\cite{Fisher67}, they are classified as first-order phase transitions (FOPTs) and continuous phase transitions. FOPTs exist more widely and obviously have a longer history and are usually described by nucleation and growth or spinodal decomposition~\cite{Gunton83,Binder,Binder2,Binder16}. Continuous phase transitions exhibit critical phenomena near their critical points~\cite{Stanley} and are characterized by universal scaling behavior. While scaling and universality in critical phenomena have been well established and accurately described by the renormalization group theory~\cite{Mask,Cardyb}, the theories for FOPTs are far less accurate~\cite{Oxtoby,Gunton,Auer,Sear,zhong18}. A way out is to see whether or not FOPTs exhibit universal scaling behavior either. However, this is a long-standing controversial issue~\cite{niehuis,fisher,binderlandau,Borgs,Marro,Bray,Rao,Lo,Dhar,Jung,Rao1,Sengupta,Mahato,Somoza,Thomas,He,Luse,zhongjcp,Hohl,zhongpret,zhongpre,zhongprl,zhongssc,kim,Suen,zhongprlc,chak,Sides98,lee,schulke,zhong02,Yildiz,zhongl05,Fan,zhong16,zhonge12,liyantao,Lee16,Pelissetto16,liange,liang,zhong18,Bar,iino,qiu,Kundu}.

\begin{figure*}
\centerline{\includegraphics[width=\linewidth]{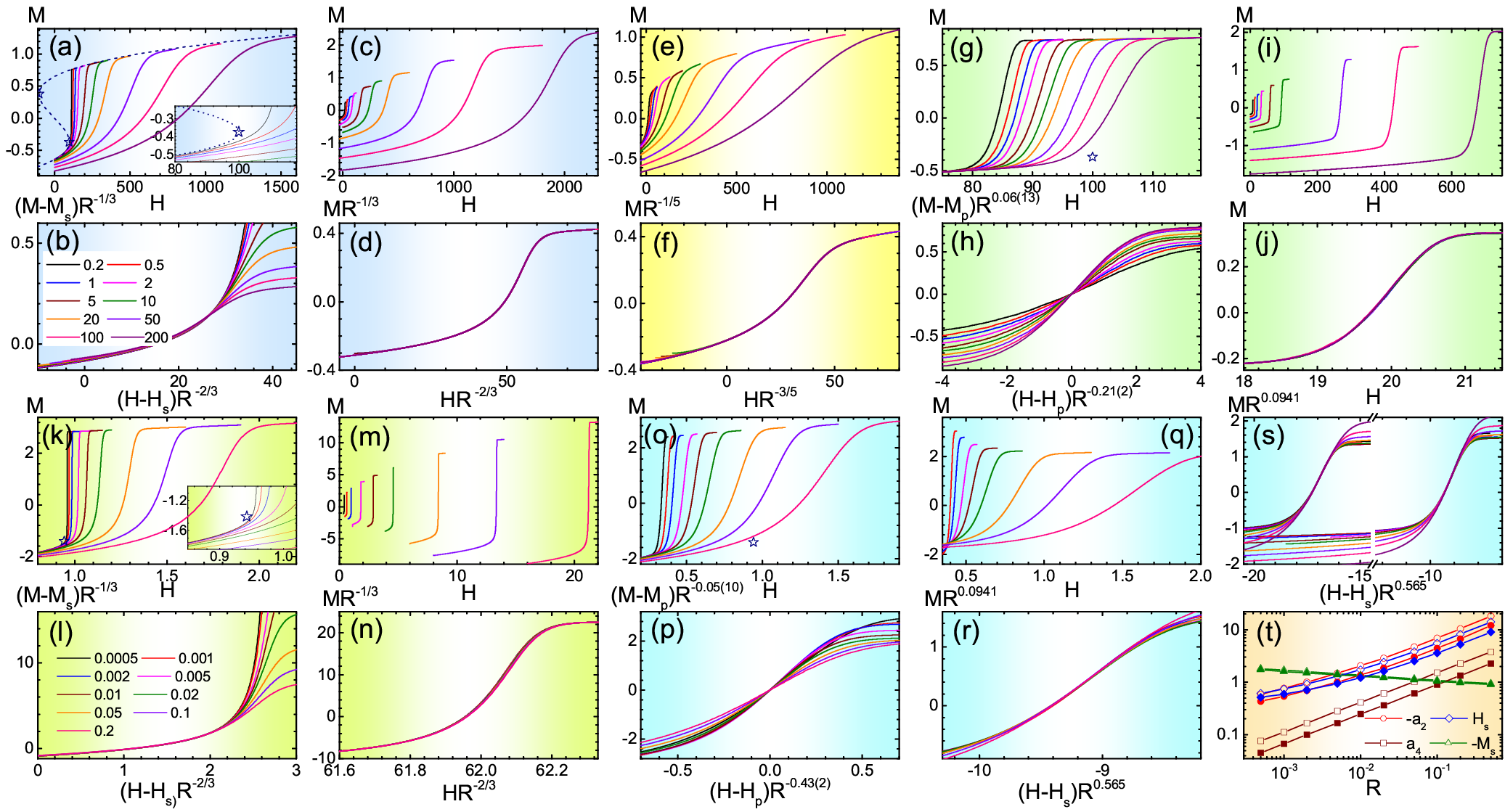}}
\caption{\label{fig}(Color online)
(a) $M$ versus $H$ for a series of rate $R$ given in (b) from numerical solutions of Eq.~(\ref{lang}). This $R$ range is adopted in (a) to (j). $a_2=a_{20}=-400$, $a_4=a_{40}=948$, $\lambda=0.000~5$~\cite{Kundu}, $\sigma=0$. The dashed curve is the mean-field equation $H=a_2M+a_4M^3$ and the stars mark the spinodal points. The inset enlarges the curves near one star. (b) Rescaling of all curves in (a) according to Eq.~(\ref{phi3}). Curves of small rates are partly cut to show the poor collapsing. (c) Similar to (a) but for $a_2=a_{20}(R_0/R)^{-1/3}$, $a_4=a_{40}(R_0/R)^{1/3}$ and $R_0=10$. (d) The same rescaling of all curves in (c) now exhibits complete collapsing. (e) Similar to (a) but with $a_2=0$, exactly the critical value. (f) Rescaling of all curves in (e) according to Eq.~(\ref{phi4}) perfectly collapses. (g) Similar to (a) but with a noise $\sigma=\sigma_0=0.000~4$. The transition occurs before the spinodal star. (h) Rescaling of all curves in (g) using the peak position $H_p$ of the fluctuations of $M$ and $M_p$ at $H_p$ together with the exponents that fit $H_p$ and $M_p$ with $R$, respectively. The exponents and their fitting standard errors are given in the axis titles. (i) Similar to (c) but with $\sigma=\sigma_0(R_0/R)^{-1}$. (j) Rescaling of all curves in (i) according to Eq.~(\ref{phi3}) exhibits complete collapsing. (k) Similar to (a) but for the two-dimensional Eq.~(\ref{lang2d}) and another series of $R$ given in (l), $a_2=a_{20}=-1$, $a_4=a_{40}=1/6$ and a small $\sigma=\sigma_0=0.000~1$ to show the universality. This $R$ range is applied to (k) to (s) and displayed in (t). (l) Rescaling of all the curves in (k) according to Eq.~(\ref{phi3}). In contrast to (b), the collapsing in the lower part is quite good because $a_4R^{1/3}$ is now small. (m) and (n) Similar to (c) and (d). Good collapses emerge again. (o) Similar to (k) but with $\sigma=\sigma_0=4$. (p) Similar to (h), collapsing occurs only at one point. (q) and (r) Similar to (o) and (p) but with $a_4=a_{40}(R_0/R)^{1/r\nu}$, $\sigma=\sigma_0(R_0/R)^{-(2\beta/\nu+z)/r}$, $R_0=0.005$, and $a_2$ given by Eq.~(\ref{a2}) and displayed in (t). $M_{s0}^2=1.9$ and $z=1.85$ and so $\beta/r\nu=0.094~1$ and $\beta\delta/r\nu=0.565$. Complete collapsing emerges but is different from (j) and (n) due to crossovers (see the text). (s) Another view of the collapsed curve in (r) (right) with $R=0.000~5$ and $R=0.5$ (purple) added and another collapsing for $R_0=0.002$ and $M_{s0}^2=2.5$ (left). (t) $R$ dependence of the parameters used in (q), (r) and (s). Open and filled symbols represent results for $R_0=0.005$ and $R_0=0.002$, respectively. Lines connecting symbols are only a guide to the eye.}
\end{figure*}

Consider a free energy~\cite{Mask}
\begin{equation}
f_4(\phi)=\frac{1}{2}a_2\phi^2+\frac{1}{4}a_4\phi^4-H\phi\label{f4}
\end{equation}
for an order parameter $\phi$ and its ordering field $H$, where $a_2$ is a reduced temperature and $a_4$ a coupling constant. It exhibits a continuous transition at $a_2=0$ and $H=0$ but first-order phase transitions for all $a_2<0$ between two ordered phases with $\phi_{\rm eq}=\pm\sqrt{-a_2/a_4}=\pm M_{\rm eq}$ at $H=0$. Dynamics can be studied by the usual Langevin equation
\begin{equation}
\frac{\partial\phi}{\partial t}=-\lambda\frac{\partial f_4}{\partial\phi}+\zeta=-\lambda\left(a_2\phi+a_4\phi^3-H\right)+\zeta,\label{lang}
\end{equation}
where $\lambda$ is a kinetic constant and $\zeta$ a Gaussian white noise whose averaged moments satisfying
$\langle\zeta\rangle=0$ and $\langle\zeta(t)\zeta(t')\rangle=2\lambda\sigma\delta(t-t')$
with a noise amplitude $\sigma$. For a field linearly changing with time $t$ with a constant rate $R$, the transition from $-M_{\rm eq}$ to $M_{\rm eq}$ does not occur at the equilibrium $H=0$. Rather, it is delayed to a larger field $H>0$. This is hysteresis that increases with the rate $R$, as seen in Fig.~\ref{fig}(a). Regarding the averaged order parameter, $M=\langle\phi\rangle$, as magnetization, this hysteresis is just the usual magnetic hysteresis. It has been continuously discovered that such a hysteresis may exhibit a power-law dependence on $R$ with a constant exponent, showing universal scaling~\cite{Rao,Lo,Dhar,Jung,Rao1,Sengupta,Somoza,Thomas,He,Luse,zhongjcp,Hohl,zhongpret,zhongpre,zhongprl,zhongssc,kim,Suen,zhongprlc,lee,schulke,zhong02,Yildiz,zhongl05,Fan,zhong16,zhonge12,liyantao,Lee16,Pelissetto16,liange,liang,zhong18,Bar,qiu}. However, even more than thirty years ago, soon after the indication of the scaling, it was found from direct numerical solutions of Eq.~(\ref{lang}) that the exponent depends instead on the amplitude $\sigma$ and thus no single power law exists~\cite{Mahato}. Recently, repeated calculations of the same model leads to a claim that the hysteresis shows power-law scaling over three orders of magnitude variation of the rate $R$ but nonuniversal exponents, which was employed to explain smaller-than-expected exponent observed in experiments and previously reported scattered exponents~\cite{Kundu}. In the following, we will demonstrate that the measured exponents are only effective. Universal exponent and even complete scaling do emerge once other effects according to an effective theory~\cite{zhongl05,zhong16} are properly taken into account. Moreover, we are able to extend this complete scaling to the more realistic two-spatial-dimensional situation in which new features appear. This confirms that FOPTs do show universal scaling described by the effective theory. It also offers a perspective different from the usual nucleation and growth~\cite{Gunton83,Binder,Binder2,Binder16} but conforming to continuous phase transitions to study FOPTs.

To start, in the absence of the noise, we note that the transition can only take place beyond the so-called spinodal point ($H_s$, $M_s$), the stars in Fig.~\ref{fig}(a), at which the free energy barrier between the two phases vanishes. As such, let~\cite{zhongl05,zhong16}
$\phi=M_s+\varphi$,
one finds $f_4(\phi)=f_4[M_s]+f_3(\varphi)+a_4\varphi^4/4$ with
\begin{equation}
f_3(\varphi)=\frac{1}{2}\tau\varphi^2+\frac{1}{3}a_3\varphi^3-h\varphi,\label{f3}
\end{equation}
where $\tau=a_2+3a_4M_s^2$, $h=H-H_s$, and $a_3=3a_4M_s$, the first two are an effective reduced temperature and an effective field, respectively, that become zero exactly at $M_s=\pm\sqrt{-a_2/3a_4}$, $H_s=a_2M_s+a_4M_s^3=2a_2M_s/3$.
One sees therefore that $\tau=0$ and $h=0$ exactly at the spinodal point, similar to $a_2=0$ and $H=0$ at the critical point. Accordingly, the transition is in fact controlled by the cubic theory $f_3$ near the spinodal where the quartic term is negligible instead of the original quartic theory to be seen below.

Next, we make a scale transformation to the dynamic equation, Eq.~(\ref{lang}), with either $f_4$ or $f_3$~\cite{zhongl05}. This is to change every quantity $O$ to $O'$ through $O=O'b^{-[O]}$ for a scaling factor $b$, where we have utilized the square brackets to denote the scale dimension. To keep the transformed dynamic equation identical with the original one, one must have
$[\lambda t]=-[a_2]$, $[H]=[\phi]+[a_2]$, $[\zeta]=[\phi]-[\lambda t]/2=[\sigma]/2$, and $[a_n]=[a_2]-(n-2)[\phi]$, representing either $a_3$ or $a_4$. Note that $\lambda$ serves only as the unit of time. In accordance with the renormalization-group theory of critical phenomena~\cite{Mask}, in mean-field theory, one sets $[a_n]=0$ even though we consider the noise. This yields
$[\phi]=[a_2]/(n-2)$, $[H]=(n-1)[a_2]/(n-2)$, and $[\sigma]=n[a_2]/(n-2)$.
Since the transformed equation is identical with the original one, they share the same solution. Accordingly, $\phi(a_2,a_n,H,\lambda t,\sigma)=b^{-[\phi]}\phi(a_2b^{[a_2]},a_nb^{[a_n]},Hb^{[h]},\lambda tb^{[\lambda t]},\sigma b^{[\sigma]})$. We then set the zero of the time at $H=0$ such that $H=R\lambda t$ (similar for $h$), replace $\lambda t$ with $R$, and let $b=R^{-1/r}$ for $r\equiv[R]$, which obeys $r=[H]-[\lambda t]=(2n-3)[a_2]/(n-2)$. These lead to the finite-time scaling (FTS) forms~\cite{Gong,Gong1,Huang},
\begin{eqnarray}
\phi=R^{1/5}g_4(a_2R^{-2/5},HR^{-3/5},\sigma R^{-4/5},a_4),\quad\label{phi4}\\
\varphi=R^{1/3}g_3(\tau R^{-1/3},hR^{-2/3},\sigma R^{-1},a_4R^{1/3},a_3),\label{phi3}
\end{eqnarray}
for the quartic and cubic theories, respectively, where $g_3$ and $g_4$ and other forms of $g$ below are universal scaling functions. Note that for the cubic theory, it is $[a_3]=[a_4]+[M_s]=0$ instead of $[a_4]$ and so $[a_4]=-[M_s]=-[\varphi]$. This is important for complete universal scaling.

The cubic FTS form can be rewritten as
$h=R^{2/3}g(\varphi R^{-1/3},\tau R^{-1/3},\sigma R^{-1},a_4R^{1/3},a_3).$
It becomes obvious that the exact mean-field exponent $2/3$ can be obtained only when all the arguments of $g$ are rate independent. In Fig.\ref{fig}(a), we plot the transition curves for a large range of $R$ for the same $a_2$, $a_4$ and $\lambda$ for $\sigma=0$. The exponent obtained by fitting $H$ at $M_s$ (so that $\varphi$ and the first argument of $g$ equal zero) with $R$ is about $0.64$ and varies with the $R$ range, slightly smaller than $2/3$. It is only effective in the sense that some other arguments of $g$ depend on $R$. The curve collapse shown in Fig.~\ref{fig}(b) is not so good indeed. However, when $a_2$ and $a_4$ properly changes with $R$ so that $\tau=a_2+3a_4M_s^2$ changes as $R^{1/3}$ according to Eq.~(\ref{phi3}), the $M$ ($=\phi$ here) vs $H$ curves in Fig.~\ref{fig}(c) perfectly collapse onto a single curve in Fig.~\ref{fig}(d). Therefore, this complete scaling in FOPTs does occur once the theory for them is fully respected.

It is essential that we do not need to subtract $M_s$ and $H_s$ in Fig.~\ref{fig}(d), because they both scale with $\varphi$ and $h$ owing to the rescaling of $a_2$ and $a_4$ with $R$ and hence just contribute rate-independent constants to an overall displacement of the rescaled curve. This seems redundant here as they are exactly known. However, it becomes important when they are not easy to be determined in the following. Note also that the curves are obtained from direct numerical solutions of Eq.~(\ref{lang}) with $a_2$ and $a_4$, not a corresponding equation with $\tau$ and $a_3$. However, the transition is controlled by the effective cubic theory with its exponents and thus follows Eq.~(\ref{phi3}) rather than the quartic theory. What does the latter do? It governs the critical behavior instead. Indeed, exactly at $a_2=0$, the critical point, the curves for a fixed $a_4$ shown in Fig.~\ref{fig}(e) then completely collapse as seen in Fig.~\ref{fig}(f) according to Eq.~(\ref{phi4}). However, excessively large $|a_2|$ will undoubtedly disrupt the complete scaling by crossovers to behaviors controlled by other fixed points.

Armed with the above method and results, it is straightforward to consider the noise. For the same $a_2$, $a_4$, $\lambda$, and $\sigma$, the exponent obtained from the fluctuation peak positions at which $HR^{-2/3}$ is a constant~\cite{Gong1} does change appreciably with $\sigma$ and the $R$ range. The original curves in Fig.~\ref{fig}(g) cannot collapse at all, see Fig.~\ref{fig}(h). However, upon scaling the noise as well as $a_2$ and $a_4$ with $R$ according to Eq.~(\ref{phi3}), the curves in Fig.~\ref{fig}(i) then completely fall onto each other in Fig.~\ref{fig}(j), exhibiting complete scaling. Therefore, the exponents that vary with $\sigma$ are only effective. The thirty-years old problem is again correctly described by the cubic theory with fully universal exponents, even if variable exponents and poor scaling collapse as exhibited in Fig.~\ref{fig}(h) are found in the usual scaling methods and may thus be considered to lack universal scaling

The important lesson learned here is that the spinodal point may be considered nonexistent since the transition now is a stochastic process and can occur before reaching $H_s$ the star in Fig.~\ref{fig}(g). This is similar to usual FOPTs that can proceed through nucleation and thus it is generally believed that there exist no spinodal curves for a system with short-ranged interactions~\cite{Gunton83,Binder,Binder2,Binder16}. However, with the present method to properly rescale the parameters so that the spinodal rescales as well, the complete scaling, including the perfect collapsing and the exponents, firmly indicates that the FOPTs do exhibit universal scaling even with noises and is indeed uniquely described by the cubic theory.

This stimulate us to apply the theory to more realistic FOPTs with spatial extension, even though the noise-activated escape problem described by Eq.~(\ref{lang}) has widespread applications~\cite{Hanggi}. We need to consider spatial fluctuations and hence the Landau-Ginzburg free-energy functional~\cite{zhongl05,zhong16}
\begin{equation}
F(\phi)=\int d^dx\left\{f_4(\phi)+\frac{1}{2}\left(\nabla\phi\right)^2\right\}\label{gl}
\end{equation}
in $d$-dimensional space. Dynamics is again governed by the Langevin equation
\begin{equation}
\frac{\partial\phi}{\partial t}=-\frac{\delta F}{\delta\phi}+\zeta=-\left(a_2\phi-\nabla^2\phi+a_4\phi^3-H\right)+\zeta,\label{lang2d}
\end{equation}
with the correlation of the noise changes to (for $\lambda=1$)
$\langle\zeta({\bf x},t)\zeta({\bf x}',t')\rangle=2\sigma\delta(t-t')\delta({\bf x}-{\bf x}')$.
To describe the FOPTs below the critical point, the lesson learned justifies our expanding again $\phi$ around a fluctuations-displaced spinodal point denoted again by $M_s$ and $H_s$ and arrive at a cubit theory, viz., replacing $\phi$ with $\varphi$ and $f_4$ with $f_3$ in Eq.~(\ref{gl}).

We still make the scale transformation with an additional $x=bx'$ for coarse graining~\cite{zhongl05,zhong16}. We now demand $F=F'$ instead of directly $[a_3]=0$ above. The results are $[t]=-[a_2]=-2$, which is set by the gradient (we then refer to Eq.~(\ref{lang}) as a zero dimension case), $[\varphi]=(d-2)/2$, $[h]=(d+2)/2$, $[a_3]=(6-d)/2$, and $[\sigma]=2[\varphi]-[t]$~\cite{note}. Accordingly, in $d=d_c=6$, the upper critical dimension, $[a_3]=0$ and the above zero-dimensional results are restored. Below $d_c$, fluctuations become important and a renormalization-group analysis parallel to critical phenomena has to be employed, giving rise to universality and scaling determined by a nontrivial fixed point. A direct consequence is that there exist anomalous dimensions such that $[t]=-z$, $[a_2]=1/\nu$, $[\varphi]=\beta/\nu$, $[h]=\beta\delta/\nu$, $[\sigma]=2\beta/\nu+z$ generally. Here, $\beta$, $\delta$, and $\nu$, and $z$ represent the exponents associated with the order parameter, the ordering field, the correlation length, and the dynamics, having identical meaning and symbols to their respective critical exponents. In $d_c$, they recover their mean-field values, $\beta=1$, $\delta=2$, $\nu=1/2$, and $z=2$. In addition, the FTS form, Eq.~(\ref{phi3}), is generalized to
\begin{equation}
M-M_s=R^{\beta\over r\nu}\hat{g}(\tau R^{- {1\over r\nu}},hR^{- {\beta\delta\over r\nu}},\sigma R^{- {\sigma\over r}},a_4R^{1\over r\nu},a_3),\label{phig}
\end{equation}
which is also identical with its critical counterpart except for the presence of $M_s$ and $H_s$, where $r=z+\beta\delta/\nu$. These demonstrate the uniformity in treating both kinds of transitions.

Apart from $M_s$ and $H_s$, the most important difference of the cubic theory to the critical theory is that its nontrivial fixed point is imaginary in value and is thus usually considered to be un-physical, though the exponents are real~\cite{zhongl05,zhong16}. Yet, counter-intuitively, it has been shown that imaginariness is physical in order for the theory to be mathematically convergent, since upon renormalization the degrees of freedom that need finite free energy costs for nucleation are integrated away so that the system becomes truly unstable at the displaced spinodal point. As a result, analytical continuation is necessary and nucleation is irrelevant to scaling~\cite{zhonge12}.

We now apply the method to Eq.~(\ref{lang2d}) in two dimensions in which $\delta=-6$ and $\nu=-5/2$ are exactly known~\cite{Cardy85} along with $\beta=1$~\cite{zhongl05,zhong16}. Before we study general cases, we choose a very small $\sigma$ such that spatial fluctuations can be ignored and the mean-field results is expected to be applicable. Figs.~\ref{fig}(k) to~\ref{fig}(m) demonstrate both the universality of the above method and the validity of our solutions. The curves for the two largest rates are slightly deviated from the main curves in Fig.~\ref{fig}(n), because their $\sigma$ is $50$ and $100$ times that of $\sigma_0$ and spatial fluctuations start to take effect. This is different from Fig.~\ref{fig}(i), where the thick curve is due to the large noise. However, when $\sigma$ gets larger, the mean-field exponents can no longer rescale the curves at all. Direct application of the above method with the two-dimensional exponents does not work either. Neither does the usual method shown in Figs.~\ref{fig}(o) and~\ref{fig}(p).

The new feature here is that there exist extra contributions with singular dimensions to the effective temperature and the spinodal field. Because $d=2<d_c$, $[a_3]=[3a_4M_s]=2$, viz., dimensional. Accordingly, $[a_4]=2-[M_s]=2-\beta/\nu=12/5$. Also, $[a_2]=1/\nu=-2/5$, $[a_4M_s^2]=8/5$, $[a_2M_s]=-4/5$, and $[a_4M_s^3]=6/5$. These indicate that the two terms both in $\tau=a_2+3a_4M_s^2$ and in $H_s=a_2M_s+a_4M_s^3$ now possess different dimensions. Moreover, the dimensions of the two terms in $H_s$ are both at odds with $[H_s]=\beta\delta/\nu=12/5$. We cannot directly multiply them by $R^{-1/r\nu}$ and $R^{-\beta\delta/r\nu}$, respectively, to achieve complete scaling as done above.

The key to solve the problem is to choose
$a_2=\tau-3a_4M_s^2-\delta a_2$ such that $a_2+3a_4M_s^2+\delta a_2$ is just $\tau$ and to subtract $H_s$. Here $\delta a_2$ denotes the fluctuation contributions from the quartic model to $a_2$. These contributions again have different dimensions for different number of loops in a standard loop expansion. A one-particle irreducible graph with $n$ vertices can be shown to have a dimension $2n/5+2$ because $[a_4]\neq4-d$ now. They are different from the cubic fluctuation contributions to $\tau$ which share identical dimensions with $\tau$ and do not need to be explicitly computed in the present approach. Besides $a_2$, quartic fluctuations do not further change $H_s$ itself due to symmetry. Cubic fluctuation contributions to $h$ also have the same dimensions and need not be considered either. Therefore, for a transition, run with $R$, to coincide with a reference curve, run using $R_0$, $a_{20}$, $a_{40}$, and $\sigma_0$, after rescaling according to Eq.~(\ref{phig}), besides directly rescaling $a_4$ and $\sigma$, we choose $a_2$ to satisfy $\tau R^{-1/r\nu}=(a_2+3a_4M_s^2+\delta a_2)R^{-1/r\nu}=(a_{20}+3a_{40}M_{s0}^2+\delta a_{20}) R_0^{-1/r\nu}$, a constant, or,
\begin{equation}
a_2=\left\{a_{20}+\delta \hat{a}_2+3a_{40}M_{s0}^2\left[1-(R_0/R)^{-{2\over r}}\right]\right\}(R_0/R)^{2\over5r}\label{a2}
\end{equation}
with $\delta \hat{a}_2=\delta a_{20}-\delta a_2(R_0/R)^{-2/5r}$, by adjusting $M_{s0}$ and $\delta \hat{a}_2$, which are determined by $a_{20}$, $a_{40}$, and $\sigma_0$ as well as $R_0$. As pointed out above, we do not need to subtract $M_s$, though it is important to deduct $H_s$, because the former shares an identical dimension with $M$ while the latter does not with $H$. In this way, although we do not need to compute $\tau_0$ and $\delta a_{20}$ and even the true $H_s$ that includes fluctuation contributions from the cubic theory, we can demonstrate through the curve collapsing whether there exists universal scaling and whether it is, if exists, described by the cubic theory.

Yet another problem is that the dynamic exponent $z$ is only known to two loops unlike the static exponents~\cite{zhong16}. We have thus to find one to best collapse the curves and thus have to estimate three free parameters $z$, $M_{s0}$, and $\delta \hat{a}_2$. Generally, given $R$ and others parameters, the slope of the curve depends on $a_2$. For $R_0=0.005$, we find $z$ cannot exceed $1.85$ in order for $R=0.01$ to fall on top of $R_0$. On the other hand, it must be bigger than $1.8$ and $1.835$ for $R=0.02$ and $R=0.05$, respectively, and the larger slightly the better. We thus choose $z=1.85$ with $M_{s0}^2=1.9(1)$ since $M_{s0}^2=2$ and $1.8$ give rise to similar collapses see Figs.~\ref{fig}(q) to~\ref{fig}(s). This $z$ is not far away from the two loop value of $1.753$~\cite{zhong16,zhong18}. The resultant $a_2$ and hence $H_s$ go smoothly with $R$ as seen in Fig.~\ref{fig}(t) although we separately adjust $\delta \hat{a}_2$ for each curve of a specific $R$, showing the consistency of the results. The same $z$ works well for $R_0=0.002$ with $M_{s0}^2=2.5(1)$, Fig.~\ref{fig}(s). Note that changing $R_0$ changes all $a_2$, $a_4$ and $\sigma$ for every $R$ and thus is drastic. One sees from Figs.~\ref{fig}(r) and~\ref{fig}(s) that good scaling collapses do emerge. The $R$ range in Fig.~\ref{fig}(r) already covers more than two orders of magnitude even though we have not shown the curve of $R=0.000~5$ as it slightly deviates. However, it follows the rule together with the largest rate $R=0.5$ shown as seen in Fig.~\ref{fig}(s), totally three orders of magnitude.

The complete scaling here for the whole transition region differs from the above ones due to crossovers as expected. These are crossovers from the metastable states [most explicitly seen from the lower curves in Fig.~\ref{fig}(s)] to the intermediate transition states and from the latter to the stable states. Both the metastable and stable states are different from the transition states and are expected to described by fixed points and exponents different from the cubic one. Crossovers between these fixed points then result in the upper and lower crossover regions that connect the intermediate transition region, which depends on $R$ like the crossover regions. Yet another crossover may occur when $a_2$ is too close to the critical point so that critical fluctuations come into play, as may be the case of $R=0.000~5$. In addition, lack of exact results for both the three parameters and the information for the crossovers leads to some uncertainty ranges in parameter selection and may deteriorate the collapse. Nonetheless, the good scaling collapses over more than two orders of magnitude firmly confirm universal scaling in FOPTs described by the cubic theory with its imaginary fixed point. In addition, unlike $a_4$ and $M_s$, $a_2$ and $H_s$ do not exhibit simple power laws with $R$ as expected, see Fig.~\ref{fig}(t), since they both consist of a series of powers with different exponents. This renders the curves in Fig.~\ref{fig}(q) different from those in Figs.~\ref{fig}(c),~\ref{fig}(i), and~\ref{fig}(m).

Finally comes the method. We solve Eq.~(\ref{lang}) by direct Euler discretization. The time step is $0.001$ for the small rates but $0.000~5$ for large ones, both having been checked to bring stable results. The average is over $10~000$ (for the large rates) to $20~000$ samples in the presence of the noise. The initial $H$ values are chosen to be sufficiently far away from the transition region so that any initial $\phi<0$ can rapidly equilibrate to the metastable state. The same discretization is also applied to solve Eq.~(\ref{lang2d}). The space step is fixed to 1, while the time step is $0.01$, again checked to be sufficient. The lattice is $240\times240$ and $480\times480$ with periodic boundary conditions. They results in curves with only negligible difference. More than $5~000$ to $10~000$ samples are employed for average.

\begin{acknowledgments}
This work was supported by National Natural Science Foundation of China (Grant No. 12175316).
\end{acknowledgments}


\begin{thebibliography}{99}
\bibitem{note} We have not considered the spatial correlation of the noise. Otherwise $[\sigma]$ is smaller by $d$ and is different from $2[\zeta]$. The resultant $[\sigma]$ has also been tested to lead to a large dynamic exponent and poorer curve collapses.
\end{thebibliography}
\end{document}